\documentclass[sigconf]{acmart}
\copyrightyear{2026}
\acmYear{2026}
\setcopyright{cc}
\setcctype{by}
\acmConference[ICSE '26]{2026 IEEE/ACM 48th International Conference on Software Engineering}{April 12--18, 2026}{Rio de Janeiro, Brazil}
\acmBooktitle{2026 IEEE/ACM 48th International Conference on Software Engineering (ICSE '26), April 12--18, 2026, Rio de Janeiro, Brazil}
\acmDOI{10.1145/3744916.3773189}
\acmISBN{979-8-4007-2025-3/26/04}
\usepackage[utf8]{inputenc}
\usepackage{color}
\usepackage{xcolor}
\usepackage{graphicx}
\usepackage{caption}
\usepackage{subcaption}
\usepackage{siunitx}
\usepackage{multirow,url,booktabs,makecell, array}
\usepackage[ruled,linesnumbered]{algorithm2e}
\usepackage{pifont}
\usepackage{bbding}
\usepackage{amsthm}
\usepackage{setspace}
\usepackage{listings}

\usepackage{enumitem}
\begin{document}


\title[LSPRAG: LSP-Guided RAG for Language-Agnostic Real-Time Unit Test Generation]{LSPRAG: LSP-Guided RAG for Language-Agnostic Real-Time \\ Unit Test Generation}

\author{Gwihwan Go}
\orcid{0009-0001-0461-9674}
\affiliation{%
  \institution{Tsinghua University}
  \city{Beijing}
  \country{China}
}
\email{iejw1914@gmail.com}

\author{Quan Zhang}
\authornote{Quan Zhang and Chijin Zhou are corresponding authors.}
\orcid{0000-0001-7778-4243}
\affiliation{%
  \institution{East China Normal University}
  \city{Shanghai}
  \country{China}
}
\email{quanzh98@gmail.com}

\author{Chijin Zhou}
\authornotemark[1]
\orcid{0000-0002-6446-247X}
\affiliation{%
  \institution{East China Normal University}
  \city{Shanghai}
  \country{China}
}
\email{tlock.chijin@gmail.com}

\author{Zhao Wei}
\orcid{0009-0007-4462-3153}
\affiliation{%
  \institution{Tencent}
  \city{Beijing}
  \country{China}
}
\email{zachwei@tencent.com}

\author{Yu Jiang}
\orcid{0000-0003-0955-503X}
\affiliation{%
  \institution{Tsinghua University}
  \city{Beijing}
  \country{China}
}
\email{jiangyu198964@126.com}

\newcommand{\tool}{\textsc{LspRag}\xspace}
\newcommand{\naive}{\textsc{\small Naive}\xspace}
\newcommand{\symprompt}{\textsc{\small SymPrompt}\xspace}
\newcommand{\draco}{\textsc{\small DraCo}\xspace}
\newcommand{\evosuite}{\textsc{\small EvoSuite}\xspace}
\newcommand{\rag}{\textsc{\small RAG}\xspace}
\newcommand{\codeqa}{\textsc{\small CodeQA}\xspace}
\newcommand{\cp}{\textsc{\small Copilot}\xspace}
\newcommand{\toolMinus}{\textsc{\small LspRag$^-$}\xspace}
\newcommand{\deepseek}{\textsc{\small DS-v3}\xspace}
\newcommand{\gptFourO}{\textsc{\small GPT4o}\xspace}
\newcommand{\gptFourOmin}{\textsc{\small GPT4o-m}\xspace}
\newcommand{\na}{-\xspace}

\newcommand{\cli}{\text{\small Commons-CLI}\xspace}
\newcommand{\csv}{\text{\small Commons-CSV}\xspace}
\newcommand{\logrus}{\text{\small Logrus}\xspace}
\newcommand{\cob}{\text{\small Cobra}\xspace}
\newcommand{\black}{\text{\small Black}\xspace}
\newcommand{\tornado}{\text{\small Tornado}\xspace}
\newcommand{\checkbox}[2][blue]{
  \tikz \node[anchor=west, scale=1, line width=1mm] 
  at (0,0) {\textcolor{#1}{\checkmark}};
}

\newcommand{\emptycheckbox}[2][gray]{
  \tikz \draw[fill=#1] (0,0) rectangle (0.6,0.6);
}

\newcommand{\coloredX}[1][red]{
  \tikz \node[anchor=west, scale=1, line width=1mm] at (0,0) {\textcolor{#1}{\ding{55}}};  
}
\newcommand{\todo}[1]{\textcolor{red}{TODO: #1}}

\newcommand{\JavaAllCov}{8.82\%\xspace} 
\newcommand{\JavaAllValid}{8.82\%\xspace}  
\newcommand{\JavaValidImpr}{8.82\%\xspace} 




\newcommand{\eg}{\emph{e.g.,}\xspace}
\newcommand{\ie}{\emph{i.e.,}\xspace}
\newcommand{\NA}{\emph{-}\xspace}

\begin{abstract}
    Automated unit test generation is essential for robust software development, yet existing approaches struggle to generalize across multiple programming languages and operate within real-time development. While Large Language Models (LLMs) offer a promising solution, their ability to generate high coverage test code depends on prompting a concise context of the focal method. Current solutions, such as Retrieval-Augmented Generation, either rely on imprecise similarity-based searches or demand the creation of costly, language-specific static analysis pipelines.
    To address this gap, we present \tool{}, a framework for concise-context retrieval tailored for real-time, language-agnostic unit test generation. \tool{} leverages off-the-shelf Language Server Protocol (LSP) back-ends to supply LLMs with precise symbol definitions and references in real time. By reusing mature LSP servers, \tool{} provides an LLM with language-aware context retrieval, requiring minimal per-language engineering effort.
    We evaluated \tool{} on open-source projects spanning Java, Go, and Python. Compared to the best performance of baselines, \tool{} increased line coverage by up to 174.55\% for Golang, 213.31\% for Java, and 31.57\% for Python. 
\end{abstract}


\maketitle
\section{Introduction}

Unit testing for modern software systems is crucial yet labor-intensive~\cite{zhao2017impact,beller2015and}. 
To ensure reliability, test suites must achieve high coverage to expose subtle bugs and edge cases. It requires developers to have a deep understanding of the internal logic and dependencies of the codebase, including how the functions under test interact with their surrounding implementation. 
This challenge is even greater in today's software development, where enterprise codebases frequently span \textit{multiple programming languages}, and developers expect \textit{real-time} test generation as they code~\cite{Abidi2019,Yang2024Multi,li2021understanding,mayer2017multi}.

Recent advances in Large Language Models (LLMs) have invigorated research on real-time multi-language unit-test generation. 
Commercial systems such as GitHub Copilot~\cite{thakkar2022copilot} already offer one-click test generation, boosting test quality with simple heuristics—for example, giving higher weight to files a developer has recently opened \cite{friedman2021copilot}. At the same time, Repository-level Retrieval-Augmented Generation~(RAG) techniques supply LLMs with additional code context obtained via textual similarity \cite{RAGsim1,RAGsim2}, graph-based relationships \cite{liu2024codexgraph,liu2024graphcoder}, or even web search \cite{zhang2024codeagent,xia2024agentless}.

\begin{figure}[t]
    \centering
    \includegraphics[width=1\linewidth]{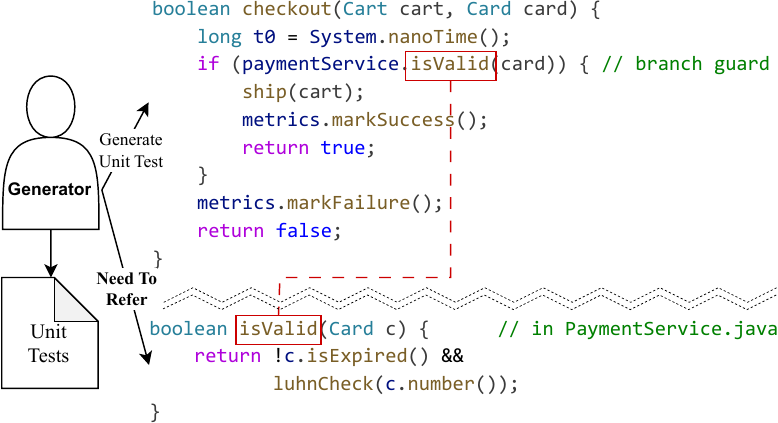}
    \caption{Unit test generation scenario for focal method \texttt{checkout}. To cover the ``true'' block of the first branch of the focal method, the generator needs the definition of \texttt{isValid} from another file.}
    \Description{Unit test generation scenario for focal method \texttt{checkout}. To cover the ``true'' block of the first branch of the focal method, the generator needs the definition of \texttt{isValid} from another file.}
    \label{fig:intoExp}
\end{figure}

{\bf Problem.}
Although existing methods enhanced code generation practice, 
they are still limited in generating high coverage unit tests, because they \textit{fail to retrieve relevant context precisely}.
For example, in Figure~\ref{fig:intoExp}, to cover the ``true'' branch of line 3 in \texttt{checkout} method, it is necessary to retrieve the definition of the guard-condition method \texttt{isValid}, which resides in another location. ( e.g., \texttt{PaymentService.java}). 
Existing approaches, however, struggle to handle this seemingly straightforward task effectively.
GitHub Copilot is not able to pull dependent context across files, so the user must supply the relevant code manually ~\cite{github2024copilottest}. 

Existing RAG approaches for code generation, unfortunately, also struggle to discover such a use-definition relationship precisely. 
This is because the RAG's embedding-and-similarity approach depends on superficial cues such as similar function names, variable names, and comments to infer relationships. These cues are often noisy and prone to variation across coding styles and conventions.
Recent research~\cite{liu2024graphcoder,DraCo} recognizes this deficiency and augments retrieval with static-analysis-based searches; however, that solution demands substantial human effort and remains tied to specific programming languages, limiting their generalizability in today's software development.  
This highlights the urgent need for a new approach for real-time and language-agnostic unit test generation. 

{\bf Insight.}
This paper introduces \tool{}, a concise context retrieval framework for multi-language unit test generation. Our key insight is that modern editors already ship with mature static analyzers exposed through the Language Server Protocol (LSP).
By querying these analyzers on demand, \tool{} obtains the precise location of context for every used symbol (e.g., function, method, etc) in the focal method.

{\bf Challenges.}
Despite having sufficient and precise symbol information, generating high-coverage unit tests remains non-trivial: there still exist challenges.
(1) \textbf{The focal method includes excessive irrelevant context.}
Achieving high test coverage requires precisely identifying the definitions of symbols that control branch conditions. 
However, in practice, a focal method often includes many symbols unrelated to the branch condition. For example, the nine-line \texttt{checkout} method in Figure~\ref{fig:intoExp} references eight external symbols (\texttt{Cart, Card, System, ..., and markFailure}), each with potentially several lines of definition. Despite this, test coverage is only influenced by the \texttt{isValid} call. While retrieving the definition for \texttt{isValid} is essential, collecting every definition within the focal method introduces unnecessary, noisy context. This excess context complicates the generator's ability to locate the correct context for solving the branch condition, ultimately hindering the generation of high-quality unit tests.

(2) \textbf{Hard to guarantee valid tests in real-time.}
Even when concise and relevant context is collected, simply providing LLMs with \textit{more context does not guarantee} the generation of valid test code. As noted in prior research~\cite{zhang2024noise,liu2024exploring,zhang2024llm}, large code context slices often introduce noise, which can obscure the original intent and reduce the LLM's ability to generate correct code. As a result, the generated test code is likely to contain syntax errors, which hinder high-coverage test generation rather than support it. Previous works~\cite{xie2023chatunitest,lemieux2023codamosa,fraser2011evosuite,pacheco2007randoop,altmayer2024coverup} have mitigated this issue by executing the generated test code in advance and collecting error messages in a fixing loop. However, this approach is not practical for a real-time scenario, as it requires a significant amount of time to compile and execute the generated test code. Furthermore, the codebase under test is often not compilable or executable in real-time.

We tackle these challenges from two dimensions.
To address Challenge (1), 
we design a key token extraction strategy that identifies symbols essential for high-coverage test generation. 
Specifically, we employ a hybrid analysis approach that combines fine-grained lexical information from the LSP with structural information from the Abstract Syntax Tree (AST), effectively filtering out unnecessary context.
To address Challenge (2), we design a compile-free self-repair mechanism that automatically fixes syntax errors in unit tests. Specifically, we utilize the LSP's diagnostic feature in a compilation-free loop and retrieve the necessary context for repairs. 
We continue this process iteratively until the errors disappear or the retry budget is exhausted, ensuring real-time performance.

We implemented \tool{} as a Visual Studio Code Extension to streamline language-agnostic unit test generation and evaluated its performance on real-world Java, Python, and Golang projects.
The results show that \tool{} consistently improves the line coverage and the rate of valid tests generated, irrespective of the programming language or the underlying LLM employed.
Our evaluation demonstrates that \tool{} yields substantial improvements in unit test quality in terms of its line coverage and valid rate. Specifically, when compared to the best value among baselines, 
\tool{} improved line coverage by a range of 91.4\% to 174.55\% for Golang projects, 27.79\% to 213.31\% for Java projects, and 16.87\% to 31.57\% for Python projects. 
Similarly, the valid rate of generated tests increased by up to 242.86\% for Golang projects, 251.91\% for Java projects, and 20.16\% for Python projects.

In summary, this paper makes the following contributions:
\begin{itemize}[leftmargin=1em]
    \item We identify a gap between academic research and industry practice: developers in large companies require high-quality test cases in real-time for different programming languages, yet concise context for LLM is difficult to obtain without compilation, execution, or heavyweight analysis.
    \item We designed and implemented \tool{}, which generate language-agnostic high-coverage unit tests in real-time. The source code is available at \url{https://thu-wingtecher.github.io/LSPRAG}.
    \item We conducted a comprehensive evaluation of \tool{} on real-world projects in three different programming languages. The results validate the ability of \tool{} to consistently improve the performance of unit test generation in terms of both line coverage and the rate of valid tests.
\end{itemize}
\begin{figure}[t]
    \centering
    \includegraphics[width=1\linewidth]{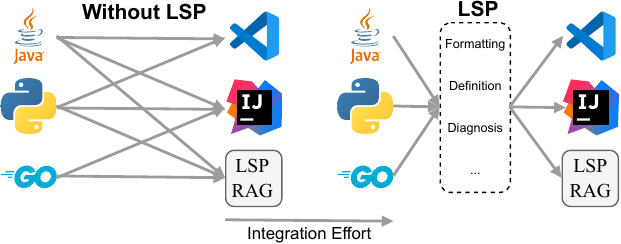}
    \caption{Integration effort before and after LSP.}
    \label{fig:lsp}
\end{figure}

\section{Background}
\subsection{Language Server Protocol}

\textit{The Language Server Protocol} (LSP) was introduced by Microsoft~\cite{microsoft2024languageprotocol} to address a significant challenge in software development: the need for each editor and IDE to have a unique implementation for every programming language's analysis features. 
As the number of editors and languages proliferated, this approach became unsustainable, forcing language authors to repeatedly rewrite the same analysis logic.
At the same time, tool vendors struggled to keep pace with the emergence of new languages. 
LSP has resolved these problems by standardizing communication between a language server and any compliant editor or IDE.

\textbf{Before vs.\ After LSP}
As illustrated in Fig~\ref{fig:lsp}, 
LSP significantly reduces implementation costs by defining a standardized communication protocol between language clients and language servers.
The language client, typically an editor or IDE used by a developer (shown on the right in each sub-figure), communicates with a standalone language server process. This server, depicted on the left, handles parsing, static analysis, and other language-specific tasks.
Before the adoption of LSP, adding rich language support required redundant effort. For instance, both Eclipse and VS Code needed separate Java plug-ins, with each language client implementing its pipeline for identical language features. LSP streamlines this process by establishing a standard protocol. Now, any compliant IDE can leverage existing language servers that also adhere to the LSP standard. This enables a single server to support multiple editors, allowing one editor to work with various languages by simply connecting to the appropriate servers.

\begin{figure*}[!ht]
    \centering
    \includegraphics[width=\textwidth]{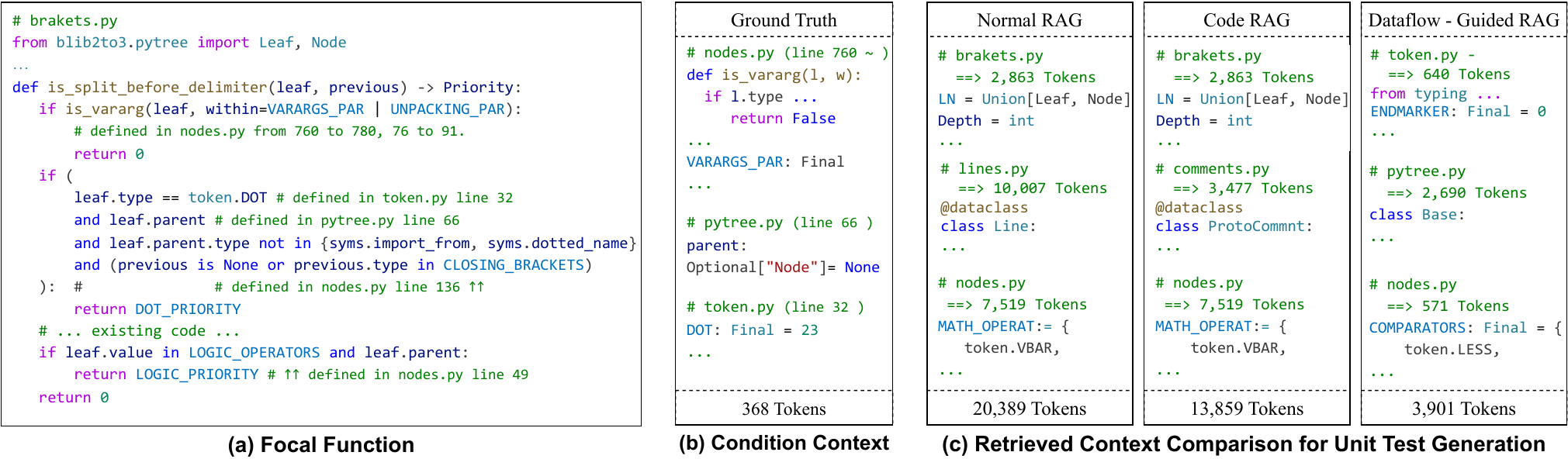}
    \caption{   
    Illustrative example of existing work's context retrieval in unit-test generation. (a) presents a simplified version of the focal method; (b) displays the ground-truth condition context and its location; and (c) compares the context slices and the number of tokens retrieved by three existing RAG techniques.}
    \label{fig:motivation}
\end{figure*}

\textbf{Typical LSP capabilities.}
A standard LSP provides a suite of powerful features, including code completion, go-to-definition, find-all-references, and real-time diagnostics that display errors and warnings as you type. It also offers hover-to-view documentation, symbol renaming, code actions for quick fixes, and refactoring. Since these capabilities are delivered through a unified protocol, users can enjoy a consistent, IDE-grade experience across any LSP-compatible editor with minimal configuration.

\subsection{Real-Time Unit Test Generation}

In today's fast-paced software development, developers frequently need to generate unit tests while writing code, especially when working on incomplete or experimental features.
The traditional approach of writing tests after the code is complete can slow down development, particularly when dealing with rapidly evolving projects.
This is because some function implementations may be incomplete or stubbed, which can temporarily leave the project not compilable and complicate test generation.
Real-time unit test generation addresses this by enabling developers to generate tests concurrently with their code, even for incomplete or partial function implementations. 
Modern tools like GitHub Copilot~\cite{thakkar2022copilot} and Amazon CodeWhisperer~\cite{amazon2024codewhisperer} leverage LLMs to create unit tests in seconds, helping developers stay in their flow. These are integrated directly into IDEs, providing test candidates within well under a minute, ensuring that the development process remains uninterrupted.

\section{Motivating Example}
Existing real-time code-generation techniques struggle to produce comprehensive unit tests, mainly because they cannot automatically retrieve the concise context that the focal method relies on.
Closing this gap is essential because exercising every branch demands a precise understanding of the symbols that guard those branches.
Without this understanding, it's difficult for LLM to understand branch conditions and fails to generate the specific inputs needed to exercise them.
To ground the problem, we present a motivating example from the project Black~\cite{black}, a widely used Python code formatter, and compare how existing research efforts attempt to retrieve necessary context.

Figure \ref{fig:motivation} (a) illustrates the simplified focal method of the Black project for which we want to generate unit tests.
The method \texttt{is\_split\_before\_delimiter} evaluates a tree leaf and its predecessor to assign a numeric priority that determines whether the formatter should split the line before that delimiter.
It skips var-arg constructs, then gives higher priority to dots in attribute chains and logical operators, returning 0 when no split is warranted.
To trigger diverse actions of this method, the generator must supply inputs that make each branch guard evaluate to true and false.
For instance, to satisfy the first guard, we must understand exactly when the function \texttt{is\_vararg} returns true and which value of leaf produces that outcome.
To do this, we need to retrieve the definition of the function \texttt{is\_vararg} from other files, which requires (1) identifying the correct file names and (2) isolating the precise code snippets that define the needed functions.

Context retrieval is a well-studied problem in the field of LLM, with many RAG techniques proposing advanced strategies.
Existing RAG techniques may address real-time unit-test generation for the focal method \texttt{is\_split\_before\_delimiter} in three main ways.
(i) Normal RAG~\cite{StandardRAG} employs similarity search with fixed-size chunking, designed for unstructured natural language.
(ii) Code-aware RAG, such as CodeRAG~\cite{Neverdecel_CodeRAG_2024} incorporates code-specific features by employing AST-based chunking and codebase indexing.
(iii) Static-analysis-based RAG, such as DraCo~\cite{DraCo}, enhances code-aware RAG by leveraging traditional program analysis techniques, such as data-flow analysis.
Although RAG techniques have shown encouraging results, they still face limitations in retrieving the necessary context for unit test generation.

These limitations are exemplified in Figure \ref{fig:motivation} (c). 
Both Normal RAG and Code RAG were able to identify only partial dependencies (e.g., \texttt{nodes.py} ) while missing other essential contexts (e.g., \texttt{pytree.py} and \texttt{tokens.py}).
This occurs because their retrieval is based on an embedding-and-similarity algorithm, which depends on textual information such as variable names, function names, and comments. 
These cues are often noisy and prone to variation across different coding styles and project conventions. 
Furthermore, they cannot isolate the relevant snippet and therefore feed the LLM large amounts of noise (e.g., 20,389 and 13,859 tokens) that hinders code generation.
As the latest open-sourced technique for static-analysis-based RAG, DraCo addressed the above limitations by building an import graph to follow dependencies, yet notable drawbacks remain.
First, it is still prone to including noise.
Because it cannot recognize which part of the imported library is used, it retrieves every imported target, thereby including unnecessary context.
For instance, it loads all constant definitions from \texttt{tokens.py} when only a single constant, \texttt{token.DOT}, is required.
Second, the implementation cost is prohibitive. Although DraCo's retrieval accuracy is relatively high, constructing and maintaining these language-specific data-flow graphs requires substantial manual effort, limiting DraCo to codebases written in Python alone.

{\bf Our approach.} Rather than hand-crafting language-specific data-flow graphs, we simply reuse the static-analysis features of LSP.
Through the LSP, these services surface the exact, minimal code fragment for requested symbols, in almost any mainstream language, giving us precise context with minimal effort. The next section details how \tool{} turns this observation into practice.
\section{Basic Concepts}\label{sec:basic_concepts}

This section formalises the four LSP providers used by \tool{}.
We introduce each concept in a top-down hierarchy—from
workspace to file, symbol, and token—so that their relationships are
clear before provider interfaces are stated.

\subsection{Workspace, File, Symbol, and Token}

\begin{definition}[Workspace]
A \emph{workspace}, denoted $\mathcal{W}$, is a finite set of
\emph{source files}:
$\mathcal{W} = \{\,f_1,\dots,f_{|\mathcal{W}|}\,\}$.
\end{definition}

\begin{definition}[File] 
Each file $f\in\mathcal{W}$ is an ordered sequence of characters and
contains a finite set of \emph{symbols},
$\Sigma_f = \{\sigma_1,\dots,\sigma_{|\Sigma_f|}\}$.
\end{definition}

\begin{definition}[Symbol]\label{def:symbol}
A \emph{symbol}~$\sigma$ is a named program entity, such as a class, function, or constant, represented by the tuple
\[
  \sigma = \bigl(\textit{name},\,\textit{kind},\,\textit{loc},\,\textit{children}\bigr).
\]
Here, \textit{name} is the identifier appearing in the source code;
\textit{kind} is an enumerated tag (e.g., \texttt{Class}, \texttt{Function}, or
\texttt{Constant}); \textit{loc} is the inclusive character location
\((f, o_{\text{start}},o_{\text{end}})\) of the symbol within its file~\(f\), start offset $o_{\text{start}}$, and end offset $o_{\text{end}}$; and
\textit{children}\(\subseteq\Sigma_f\) is the set of nested symbols whose
spans lie strictly inside~\textit{loc} (e.g., a method symbol is one of the children from its class symbol.)
\end{definition}

\begin{definition}[Token]\label{def:token}
A \emph{token}~$\tau$ is the smallest lexical unit recognised through LSP. 
We describe $\tau$ as the tuple
\[
  \tau = \bigl(\textit{loc},\,\textit{lex},\,\textit{tok\_kind}\bigr),
  \qquad f\in\mathcal{W}.
\]
The \(f\) denotes the file containing the token;
\textit{loc} is the character location \((f, o_{\text{start}},o_{\text{end}})\);
\textit{lex} is the raw lexeme text; and \textit{tok\_kind} indicates the lexical
category (e.g., identifier, keyword, \dots). 
\end{definition}

\subsection{Provider Interfaces}
In addition to the above concepts, we specify the four code-context providers implemented on top of LSP that \tool{} invokes.
\begin{definition}[Symbol Provider]
The \emph{Symbol Provider}, denoted $\textit{SYM}$, is a function that, given a file $f \in \mathcal{W}$, returns the set of all symbols in $f$ that are not nested within another symbol.
\[
  \textit{SYM}(f) = \Sigma_f \setminus \bigcup_{\sigma' \in \Sigma_f} \sigma'.\textit{children}
\]
where $\Sigma_f$ is the set of all symbols in file $f$.
\end{definition}

\begin{definition}[Token Provider]\label{def:stk}
The \emph{token provider}, denoted $\textit{TOK}$, is a function that, given a location $loc_q \in \mathcal{W}$, returns an ordered sequence of all tokens whose spans are contained within $loc_q$.
Let $T_f$ be the complete, ordered sequence of all tokens in file $f$. The function is defined as:
\[
  \textit{TOK}\bigl(loc_q\bigr) = \langle \tau \in T_f \mid \tau.\textit{loc} \subseteq loc_q \rangle
\]
The output is a sequence, not a set, preserving the original order of the tokens as they appear in the file.
\end{definition}

\begin{definition}[Definition Provider]\label{def:def}
  A \emph{Definition Provider}, denoted $\textit{DEF}$, is a function that maps a given token~$\tau$ to the set of locations~$L_\tau$ of its corresponding definition symbols. This can be expressed as:
  \[
    \textit{DEF}(\tau) = L_\tau \triangleq
    \begin{cases}
      \{ \sigma.\textit{loc} \mid \sigma \text{ is a definition for } \tau \} & \text{if definitions exist} \\
      \emptyset & \text{otherwise}
    \end{cases}
  \]

  Subsequent to the definition retrieval, resolving a location~$l$ to a symbol~$\sigma$ involves finding the symbol within the file's symbol set from $\textit{SYM}(l.f)$ that satisfies:
  \[
  \sigma \in \textit{SYM}(l.f) \land \sigma.\textit{loc} = l
  \]
  Note that the file containing the definition, $l.f$, may reside outside of the current workspace~$\mathcal{W}$, for instance, within an external library or a standard system header.
\end{definition}

\begin{definition}[Reference Provider]\label{def:ref}
The \emph{reference provider}, denoted $\textit{REF}$, performs the inverse operation of the
definition provider. Given a token $\tau$, it finds all tokens in the workspace that refer
to the same definition. Let $\sigma_{\textit{def}}$ indicates the definition symbol for the token $\tau$. The function is defined as:
\[
  \textit{REF}(\tau) =
  \begin{cases}
    \begin{aligned}
      \{ \tau'.\textit{loc} \mid{} & \tau' \in \bigcup_{f \in W} T_f \\
                                  & \land \textit{DEF}(\tau') = \sigma_{\textit{def}} \}
    \end{aligned}
    & \text{if } \sigma_{\textit{def}} \neq \texttt{null} \\
    \emptyset & \text{if } \sigma_{\textit{def}} = \texttt{null}
  \end{cases}
\]
The output is a set of locations, each corresponding to a token that references the same definition symbol.
\end{definition}

\section{Design of \tool{}}
\label{sec:design}
\begin{figure}[t]
    \centering
    \includegraphics[width=1\linewidth]{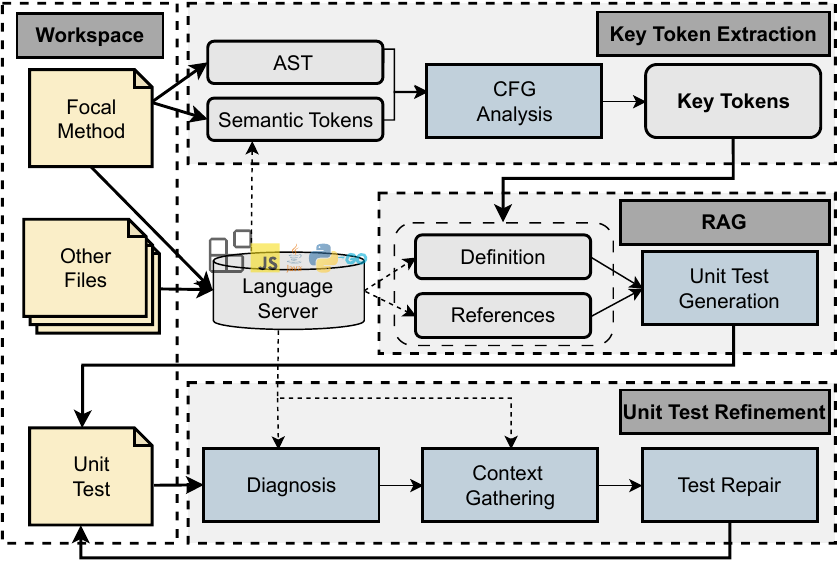}
    \caption{Overall workflow of \tool{}.}
    \label{fig:overview}
\end{figure}

This section presents the design of \tool{}, an enhanced RAG framework for multi-language real-time unit test generation.
As illustrated in Figure~\ref{fig:overview}, \tool{} consists of three core modules that operate on a developer's workspace $\mathcal{W}$.
We assume that a developer is working on a focal method within a file $f \in \mathcal{W}$, which may have dependencies on various symbols defined in other files.

Upon a developer's request to generate a unit test, \tool{} begins with \textbf{Key Token Extraction} module.
This module is responsible for identifying a set of key tokens within the focal method. These tokens are pivotal in guiding the generation of high-coverage test cases as they typically govern the method's control flow and its interactions with external components.
Subsequently, these extracted tokens are passed to the \textbf{RAG} Module. This module leverages \textit{DEF} and \textit{REF} providers to gather contextual information, including definitions and usages of the key tokens. This context is then used to construct a detailed prompt that is fed to an LLM.
Finally, to address potential syntactic errors in the generated test code, the \textbf{Unit Test Refinement} module inspects the test code without compilation. If any potential errors are detected, it gathers the necessary context for remediation and feeds this information back to the LLM to correct the code.

\subsection{Key Token Extraction}
\label{sec:key_token_extraction}

\begin{figure}[t]
    \centering
    \includegraphics[width=1\linewidth]{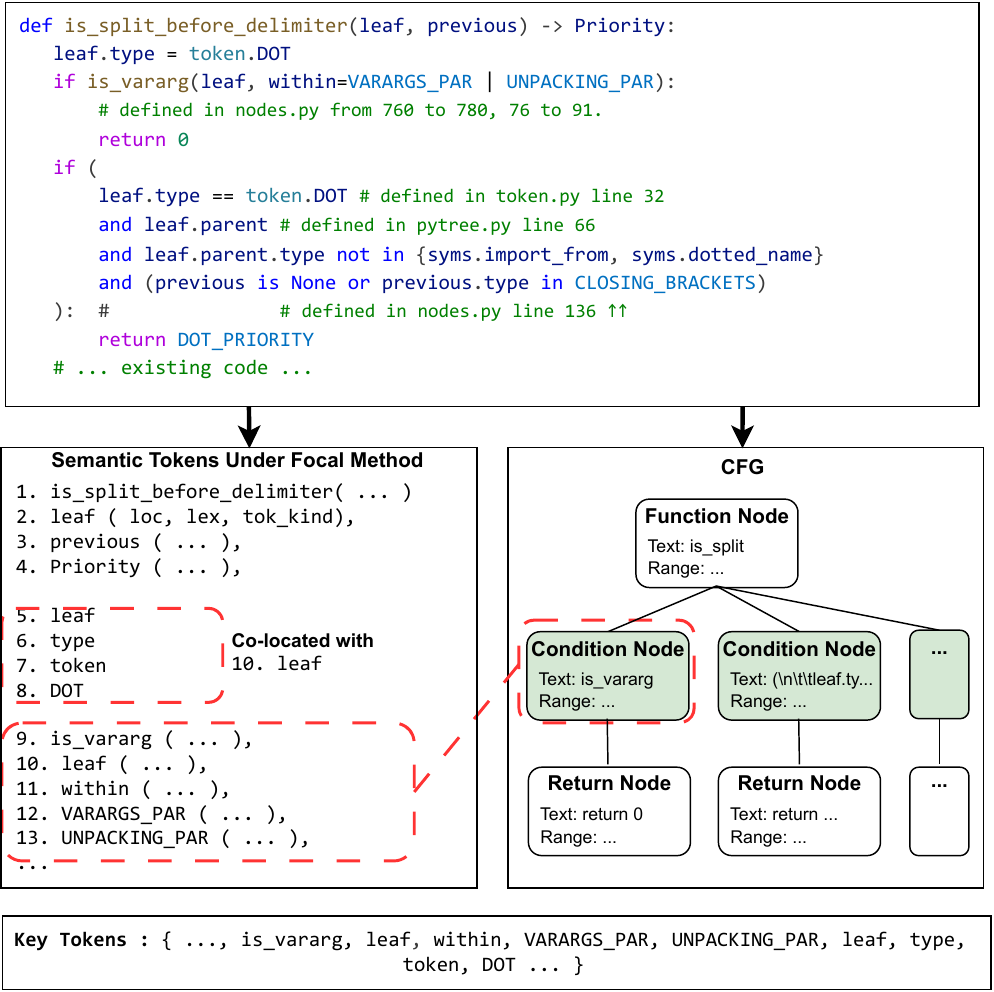}
    \caption{This figure illustrates the key token selection process. The red box highlights how key tokens are identified. Using the lexical information and CFG, \tool{} identifies those tokens involved in or influencing branch decisions.}
    \label{fig:tokenExtraction}
\end{figure}

A primary challenge in context retrieval is that a focal method often contains many tokens irrelevant to its core logic. Including context of these tokens can degrade the quality of the generated tests by retrieving non-essential context.
To address this, we introduce the concept of a \emph{key token}, which is a token~$\tau$ whose semantics are essential for constructing high-coverage tests.
Specifically, we define key tokens as those that represent control-flow decisions or external dependencies.
To identify these tokens, \tool{} performs a hybrid analysis that combines fine-grained lexical information with a structural understanding, as illustrated in Figure~\ref{fig:tokenExtraction}.

\subsubsection{Lexical Information from LSP.}

\tool{} begins by invoking a token provider, \textit{TOK} (Definition~\ref{def:stk}), to retrieve a sequence of all tokens within the scope of the focal method's location, $loc_m$.
\[
  \langle \tau_1, \tau_2, \dots, \tau_n \rangle = \textit{TOK}(loc_m)
\] 
Each token $\tau$ in this sequence includes its location, lexical name, and a semantic role, such as \texttt{parameter} or \texttt{identifier}. This information is necessary for the subsequent context retrieval phase, because the context provider on top of LSP accepts input as a specific location of a token. Additionally, based on semantic role, one can filter out unnecessary tokens, such as constants.
While this lexical information is useful, it is insufficient on its own because it lacks structural context.
For example, tokens like ``if'' and ``def'' are both classified simply as ``Keyword'' by \textit{TOK}, despite their different structural implications for control flow. 
The LSP-oriented solution cannot resolve this limitation, since it is designed for editor-centric features, not compiler-level control-flow analysis\footnote{\url{https://github.com/microsoft/language-server-protocol/issues/1675}}. 
Consequently, relying solely on LSP features is insufficient for identifying the key tokens needed for high-coverage test generation.

\subsubsection{Structural Analysis via AST.}
To overcome the limitations of the LSP, \tool{} leverages the Abstract Syntax Tree (AST). 
The AST provides a language-agnostic structural view that is essential for understanding program hierarchy and can be easily implemented in a language-agnostic way, which is crucial for multi-language support.
The effectiveness of this technique has been demonstrated in prior RAG research~\cite{DraCo, code_qa}.
\tool{} reuses widely-used Tree-sitter~\cite{tool:treesitter} to construct a AST for focal method. The critical step in this process is to bridge the structural AST view with the lexical LSP tokens. To achieve this, \tool{} preserves the precise source code location (i.e., line and column numbers) of every node during AST construction. This allows for a precise mapping between an AST node and its corresponding set of tokens if the node's location span contains the tokens' locations.

\subsubsection{Key Token Identification.}

With the combined data, the tool constructs a lightweight, language-agnostic Control-Flow Graph (CFG) for the focal method. The CFG enables path-sensitive analysis to identify the final set of key tokens, $T_{\text{key}}$. At a high level, a token~$\tau$ from the focal method is added to $T_{\text{key}}$ if it participates in or affects the change of a conditional expression that determines a branch in the CFG. Specifically, we first identify tokens $\tau'$ that participate in a conditional expression. Next, we filter out the token sequence $\textit{TOK}(loc_m)$ that does not need a context search by excluding tokens with semantic roles such as \texttt{Keyword}, \texttt{Identifier}, \texttt{Literal}, \texttt{Comment}, \texttt{String}, and \texttt{Regex}. In general, these tokens are either constants or syntax keywords that do not require a context search. Then, we extract tokens from $\textit{TOK}(loc_m)$ that appear on the same line as the token $\tau'$ and add them to $T_{\text{key}}$.
The resulting set $T_{\text{key}}$ represents the tokens most likely to affect the focal method's execution path. It is then passed to the subsequent RAG module.

\subsection{RAG}
\label{sec:rag}

This module is designed for precise context retrieval for the set of key tokens, $T_{\text{key}}$, identified in the previous stage. To achieve this, \tool{} systematically queries the LSP's \textit{DEF} (Definition~\ref{def:def}) and \textit{REF} (Definition~\ref{def:ref}) for each token in $T_{\text{key}}$ and assembles the retrieved information into a coherent prompt.

\subsubsection{Context Retrieval.}
To retrieve a token's definition, \tool{} invokes the \textit{DEF} to find the location where the token is defined, which we denote as $loc_{\text{def}}$. By default, \textit{DEF} resolves symbols across the entire project space, including standard and third-party libraries.
To maintain contextual relevance, \tool{} filters these results, considering only definitions within the current workspace $\mathcal{W}$. This prevents the inclusion of unnecessary definitions from standard libraries (e.g., String, Integer). This filtering is expressed as:
\[
\text{process } loc_{\text{def}} \text{ only if } loc_{\text{def}}.f \in \mathcal{W}, \text{ where } loc_{\text{def}} = \textit{DEF}(\tau.loc)
\]
If a valid definition location is found, \tool{} first locates $\sigma_{\text{def}}$ at $loc_{\text{def}}$ and then extracts the source code from $\sigma_{\text{def}}.loc$.

Next, to find usage examples, \tool{} queries the \textit{REF} to find all use-sites of the token's symbol within the workspace~$\mathcal{W}$.
We denote $L_{\text{refs}}$ as the queried result of $\textit{REF}(\tau)$.
A raw reference location $loc_r' \in L_{\text{refs}}$ (e.g., a variable in an expression) is often insufficient, as it only points to the token at other file without its surrounding context. To create a meaningful example, \tool{} enriches each reference. For each location $loc_r' \in L_{\text{refs}}$, the tool identifies the smallest symbol $\sigma_{\text{enclosing}}$ that fully contains the reference location ($loc_r' \subseteq \sigma_{\text{enclosing}}.\textit{loc}$). This is achieved by retrieving all top-level symbols in the reference's file $loc_r'.f$ through $\textit{SYM}$ and then recursively searching their children. The source code of $\sigma_{\text{enclosing}}$ is then extracted, transforming a simple location into a complete usage example (e.g., the entire function that uses the token $\tau$), which demonstrates how the symbol is used in other locations.

\subsubsection{Prompt Construction.}
The final step synthesizes the retrieved information into a structured prompt for the LLM. This prompt is designed to guide the LLM in generating comprehensive unit tests and consists of three parts:
(1) The full source code of the focal method to be tested.
(2) The definitions and references of the key tokens retrieved previously.
(3) A lightweight unit-test template with necessary import statements and class or function structures inferred from the focal method's file.
This structured prompt provides the LLM with focused and relevant information, minimizing noise that can degrade code generation quality and empowering it to produce accurate and thorough unit tests.
Due to page limitations, we provide the full prompt format in our artifact, which is open-sourced. The prompt is designed using CoT~\cite{wei2022chainofthought} with a one-shot setting~\cite{brown2020language}. 

\subsection{Unit Test Refinement}

While LLMs excel at code generation, their output can be imperfect, containing syntactic errors that cause compilation failures. To address this, \tool{} employs an automated refinement module that detects and repairs them. This process operates in two phases: (1) real-time error detection and (2) context-aware error repair.

\subsubsection{Real-Time Error Detection.}
\tool{} leverages the diagnostic capabilities of the LSP to perform immediate analysis on the generated code. 
In detail, when code is modified within a file $f \in W$, \tool{} notifies the corresponding language server. 
The server, in turn, analyzes the file and returns a set of diagnostics. 
Each diagnostic is a tuple $(m, loc)$, where $m$ is a human-readable error message (e.g., ``undefined variable 'x''') and $loc$ is the location of the code that triggers the error. 
This mechanism provides instantaneous feedback without requiring a project compilation or code execution, which is critical for real-time context.

\subsubsection{Context-Aware Error Repair.}
\begin{table}[ht!]
\centering
\caption{Error Categories Grouped by Necessary Context}
\label{tab:error_context_groups}
\resizebox{\linewidth}{!}{
\begin{tabular}{p{4.6cm} S[table-format=5.0] p{4.5cm}}
\toprule
\textbf{Error Category} & {\textbf{Frequency}} & \textbf{Context Needed for Fix} \\
\midrule

Redeclaration/Duplicate Definition & 28300 & \multirow{2}{4.5cm}{Workspace-level context

(e.g., project structure)} \\
Import/Module Resolution Error     & 13517 & \\
\midrule

Member Access/Usage Error  & 13387 & \multirow{3}{4.5cm}{Symbol-level context 

(e.g., context of a specific symbol)} \\
Type Mismatch/Compatibility Error & 4388  & \\
Constructor Call Error            & 1670  & \\
\midrule

Syntax Error                      & 5467  & \multirow{2}{4.5cm}{No external context required} \\
Unhandled Exception               & 179   & \\
\midrule

\textbf{Total} & \textbf{66908} & \\
\bottomrule
\end{tabular}
}
\end{table}
Upon receiving a set of non-empty diagnostics, \tool{} initiates the repair phase. 
The core challenge in automated code repair is providing the LLM with sufficient and relevant context to understand and fix the error. 

To determine the optimal context for different error types, 
We constructed the dataset (Table~\ref{tab:error_context_groups}) ourselves by collecting ~70k errors across Java, Python, and Golang, since no dataset exists for LLM-generated code paired with error messages. These errors were generated by prompting diverse LLMs to produce test cases for different projects (including those used in evaluation), and then diagnosing them via the LSP-based workflow. After collection, we manually analyzed the errors, identifying their root causes and the contextual information helpful in fixing them. This information is general rather than tied to specific projects. For instance, “imported and not used” consistently indicates import-related issues, while “redeclared in this block” points to symbol redeclaration errors.
Based on the analysis result, \tool{} gathers the corresponding necessary context by invoking the providers implemented on top of LSP.
For a given diagnostic $(m, loc_{err})$, \tool{} collects context as follows: 
\begin{itemize}[leftmargin=1em]
    \item \textbf{Symbol-Level Context:} 
    For errors related to a specific symbol (e.g., undefined variables, incorrect method calls), \tool{} retrieves its definition and usage examples. It invokes \textit{DEF} to fetch the canonical definition and \textit{REF} to find usage examples for the tokens at $loc_{err}$, providing the LLM with both the ground-truth specification and concrete examples of correct usage.
    \item \textbf{Workspace-Level Context:} 
    For errors involving project structure 
    (e.g., missing imports), 
    a broader view of the project is necessary. \tool{} provides this by supplying: (1) the workspace's file structure and (2) a list of top-level symbols (e.g., classes, functions) in the file where the error occurred ($loc_{err}.f$).
\end{itemize}

Finally, \tool{} constructs a prompt containing the erroneous code, the diagnostic message~$m$, and the retrieved context.
This process is iterative: after the LLM proposes a fix, \tool{} re-triggers the diagnostic mechanism. 
This loop continues until all diagnostics are resolved or a predefined number of attempts is reached.
\section{Implementation}
We implemented a prototype of \tool{} in TypeScript, comprising approximately 22k lines of code out of 8k lines of unit test code. 
The implementation was guided by two primary goals: developer usability and language generalizability.

To enhance usability, \tool{} was implemented as a VS Code extension, which integrates seamlessly into the developer's existing workflow and reduces the barriers to adoption. The tool leverages the VS Code Extension API, which provides comprehensive capabilities for user interface interactions, editor functionality, and command handling. 
To achieve language generalizability, we designed \tool{} to be language-agnostic. It leverages the LSP to consume semantic information from any compliant language server and utilizes Tree-sitter \cite{tool:treesitter} for robust source code parsing into an AST. Consequently, \tool{} is compatible with any programming language supported by an LSP server and Tree-sitter. We have validated \tool{}'s functionality with several major language servers, including Pylance for Python \cite{vscode:python}, the Oracle Extension Pack for Java \cite{vscode:java}, and the official Go extension \cite{vscode:go}.

\tool{} currently offers two main features for unit test generation. The first is method-level test generation, which automatically generates a unit test class for a given focal method. Such a class contains multiple test functions, each designed for a unique execution branch within the focal method. The second feature, branch-targeted test generation, allows developers to select a specific conditional branch within a focal method and generates a set of test cases specifically designed to exercise the logic of that branch.
\section{Evaluation}
\label{sec:5_eval_sec}
In this section, we comprehensively evaluate \tool{}'s performance on real-world projects across different programming languages. Our evaluation investigates the following questions:

\begin{itemize}[leftmargin=*]
  \item \textbf{RQ1}(\S7.2): Can \tool{} generate \emph{higher coverage} unit tests than other baselines?
  \item \textbf{RQ2}(\S7.3): Does \tool{} maintain low latency and efficient token usage during generation?
  \item \textbf{RQ3}(\S7.4): How much does each pipeline component of \tool{} contribute to test coverage?
\end{itemize}
\subsection{Experiment Setup}
{\bf Benchmarks.} 
Our evaluation is conducted on six open-source projects written in Python, Java, and Golang. 
Python and Java were selected due to their common use in prior unit test generation research. 
Golang was chosen as it is widely adopted in industry but less examined in academic studies, which we anticipated would present different challenges for LLMs.
As shown in Table~\ref{tab:dataset-stats}, we selected two projects per language, all of which are frequently used as benchmarks in the unit test generation domain~\cite{xie2023chatunitest,wang2024hits,NxtUnitGo,lemieux2023codamosa,lukasczyk2022pynguin}.
\begin{table}[ht!]
    \centering
    \caption{Dataset Statistics}
    \label{tab:dataset-stats}
    \resizebox{\linewidth}{!}{
    \begin{tabular}{l|l|c|c}
    \toprule
    \textbf{Project}                  & \textbf{Domain}            & \textbf{Version} & \textbf{Language} \\ \midrule
    Black~\cite{black}                          & Code formatter      & 8dc9127          & Python                 \\
    Tornado~\cite{tornado}                  &  Network/Web framework   & 81d36df1          & Python                 \\
    Commons-CLI~\cite{commons-cli}                & Command line interface parser       & 266ab84a           & Java                 \\
    Commons-CSV~\cite{commons-csv}            & Library for processing CSV files       & ca3a95c3         & Java                 \\
    Cobra~\cite{spf13_cobra}                  & Framework for creating Go CLI       & ceb39ab          & Golang                 \\
    Logrus~\cite{logrus}                      & Structured logging library           & d1e6332          & Golang                 \\
    \bottomrule
    \end{tabular}
    }
    \end{table}

\begin{table*}[t]
    \centering
    \small
    \caption{Comparison of line coverage and valid rate across all baselines. Our approach, \tool{}, consistently outperforms all baselines, achieving higher scores across diverse projects and models. Missing values (--) indicate that the baseline does not support test generation for those projects.}
    \label{tab:main_results}
    \resizebox{\linewidth}{!}{
    \begin{tabular}{@{} cl *{5}{S[table-format=1.2]} *{5}{S[table-format=1.2]} @{}}
      \toprule
      \multirow{2.5}{*}{\textbf{Project}} & \multirow{2.5}{*}{\textbf{Model}} & \multicolumn{5}{c}{\textbf{Coverage}} & \multicolumn{5}{c}{\textbf{Valid Rate}} \\
      \cmidrule(lr){3-7} \cmidrule(lr){8-12}
      & & {\codeqa} & {\rag} & {\symprompt} & {\textbf{\tool{}}} & {\draco} & {\codeqa} & {\rag} & {\symprompt} & {\textbf{\tool{}}} & {\draco} \\
      \midrule
      \multirow{3}{*}{\black{}} 
      & \gptFourOmin{}  & 34.92\% & 26.29\% & 22.09\% & \textbf{41.91\%} & 32.58\% & 55.57\% & 37.26\% & 66.35\% & \textbf{79.73\%} & 54.52\% \\
      & \gptFourO{}     & 40.47\% & 29.94\% & 23.99\% & \textbf{48.10\%} & 35.33\% & 79.46\% & 77.53\% & 59.87\% & \textbf{85.15\%} & 73.31\% \\
      & \deepseek{}     & 46.03\% & 38.98\% & 34.18\% & \textbf{56.89\%} & 37.64\% & 59.26\% & 72.24\% & 75.18\% & \textbf{88.03\%} & 71.04\%  \\
      \midrule 
      \multirow{3}{*}{\tornado{}}
      & \gptFourOmin{}  & 29.04\% & 49.44\% & 31.44\% & \textbf{57.78\%} & 46.54\% & 71.26\% & 64.30\% & 77.01\% & \textbf{81.84\%} & 75.82\% \\
      & \gptFourO{}     & 29.97\% & 46.29\% & 34.35\% & \textbf{60.90\%} & 41.09\% & 80.14\% & 83.95\% & 72.82\% & 84.53\% & \textbf{85.03\%}  \\
      & \deepseek{}     & 38.45\% & 54.05\% & 50.78\% & \textbf{64.62\%} & 48.99\% & 85.93\% & 81.84\% & 85.30\% & \textbf{91.13\%} & 90.48\% \\
      \midrule
      \multirow{3}{*}{
      \parbox[c]{0.85cm}{\centering Commons \\ CLI}
      }
      & \gptFourOmin{}  & 10.63\% & 05.02\% & 02.76\% & \textbf{33.29\%} & {--} & 12.40\% & 08.18\% & 07.05\% & \textbf{45.08\%} & {--} \\
      & \gptFourO{}     & 09.60\% & 03.21\% & 03.00\% & \textbf{34.69\%} & {--} & 08.23\% & 07.27\% & 07.20\% & \textbf{48.12\%} & {--} \\
      & \deepseek{}     & 20.72\% & 17.68\% & 05.62\% & \textbf{37.72\%} & {--} & 13.29\% & 14.55\% & 09.21\% & \textbf{60.52\%} & {--} \\
      \midrule
      \multirow{3}{*}{
      \parbox[c]{0.85cm}{\centering Commons \\ CSV}
      }
      & \gptFourOmin{}  & 40.97\% & 24.93\% & 18.50\% & \textbf{80.51\%} & {--} & 23.64\% & 13.20\% & 06.28\% & \textbf{82.85\%} & {--}   \\
      & \gptFourO{}     & 44.85\% & 44.84\% & 25.28\% & \textbf{78.33\%} & {--} & 20.68\% & 26.53\% & 14.41\% & \textbf{90.90\%} & {--}  \\
      & \deepseek{}     & 65.11\% & 44.68\% & 35.07\% & \textbf{83.20\%} & {--} & 43.27\% & 32.24\% & 29.82\% & \textbf{90.95\%} & {--}  \\
      \midrule
      \multirow{3}{*}{\cob}
      & \gptFourOmin{}  & 07.11\% & 12.03\% & 03.39\% & \textbf{23.03\%} & {--} & 06.01\% & 09.50\% & 01.23\% & \textbf{23.88\%} & {--} \\
      & \gptFourO{}     & 10.05\% & 07.57\% & 00.22\% & \textbf{27.60\%} & {--} & 09.70\% & 08.12\% & 00.81\% & \textbf{33.27\%} & {--} \\
      & \deepseek{}     & 15.50\% & 13.01\% & 08.57\% & \textbf{37.22\%} & {--} & 10.30\% & 10.69\% & 02.78\% & \textbf{34.65\%} & {--} \\
      \midrule
      \multirow{3}{*}{\logrus}
      & \gptFourOmin{}  & 05.52\% & 11.14\% & 00.23\% & \textbf{23.71\%} & {--} & 14.32\% & 20.83\% & 00.83\% & \textbf{34.02\%} & {--} \\
      & \gptFourO{}     & 05.61\% & 13.09\% & 00.23\% & \textbf{27.75\%} & {--} & 14.17\% & 26.52\% & 00.83\% & \textbf{32.02\%} & {--} \\
      & \deepseek{}     & 11.34\% & 11.00\% & 05.43\% & \textbf{21.81\%} & {--} & 13.33\% & 15.83\% & 07.5\% & \textbf{33.11\%} & {--} \\
      \bottomrule
    \end{tabular}
    }
  \end{table*}

{\bf Baseline Selection.}
To evaluate the performance of \tool{}, we selected four baselines, comprising three RAG techniques and one prompt engineering method.
\begin{itemize}[leftmargin=*]
  \item \rag{}: A normal RAG with OpenAI embeddings and FAISS index via LangChain. It retrieves the top 3 most similar snippets based on cosine vector similarity.
  \item \codeqa{}~\cite{code_qa}: An open-source RAG pipeline designed for code generation. Its approach, which involves indexing the codebase and parsing code structure with an AST, is analogous to that used in modern AI-assisted IDEs like Cursor~\cite{cursoride}.
  \item \draco{}~\cite{DraCo}: A RAG pipeline integrated with program analysis. Although it is specific to Python, this tool's use of static analysis-guided retrieval plays a comparable baseline for evaluating the context retrieval capabilities of \tool{}.
  \item \symprompt{}~\cite{symprompt}: A prompt engineering technique for unit test generation. It proposes a novel prompt format that translates CFG paths into descriptive comments, thereby improving the readability and structure of the generated tests. We replicated this prompt format using our own CFG.
\end{itemize}

{\bf Model Selection and Environment.}
Our evaluation utilizes three LLMs to assess \tool{}'s performance across different architectures and scales. 
We selected GPT-4o~\cite{openai2024gpt4technicalreport} and its smaller variant, GPT-4o-mini (\gptFourOmin{}), to represent the transformer architecture. To test against a different model design, we also included Deepseek-V3~\cite{liu2024deepseek} (\deepseek{}), which is based on a Mixture-of-Experts architecture. 
The experiments were conducted on a server running Ubuntu 22.04 LTS, equipped with a 128-core AMD EPYC 7763 CPU and an NVIDIA V100 GPU featuring 32GB of memory. For all language models, we used the default generation parameters, including temperature, to ensure a consistent baseline.

\subsection{Generation of High Coverage Unit Tests}
\label{sec:evrq1}
To answer RQ1, we evaluated \tool{}'s ability to generate high coverage unit tests against selected baselines. We measured performance using two key metrics: line coverage to quantify test effectiveness, and valid rate to assess the proportion of syntactically correct, usable tests. We define the valid rate as the ratio of grammatically correct test cases to the total number of unit test generation attempts. Since invalid test cases are not compilable or executable, they cannot produce any coverage.

{\bf Evaluation Setup.} 
For each project listed in Table~\ref{tab:dataset-stats}, we generated unit tests for methods exceeding 10 lines of code, as these typically contain complex logic that is difficult to fully cover. This resulted in 299 focal methods for Black, 521 for Tornado, 101 for Cobra, 24 for Logrus, 43 for Commons-CLI, and 145 for Commons-CSV. To ensure reliable results, we repeated each experiment five times and report the average values. We configured \tool{} to perform a maximum of five self-correction iterations.

{\bf Statistical Analysis.}
To verify the significance of our findings, we performed statistical tests on the results. We computed p-values to assess the significance of the improvements and reported effect sizes to quantify their magnitude, following the guidelines from Arcuri et al~\cite{arcuri2014hitchhiker}. The null hypothesis ($H_0$) posits that there is no statistically significant difference in line coverage between \tool{} and the baselines. Conversely, the alternative hypothesis ($H_1$) states that \tool{} achieves a statistically significant improvement.

{\bf Results.}
The overall results, presented in Table~\ref{tab:main_results}, show that \tool{} consistently improves line coverage across all projects, regardless of the programming language or the underlying LLM. Our statistical analysis confirms that these improvements are significant, with all p-values being less than 0.05, thus supporting the $H_1$. This finding highlights the effectiveness of \tool{}'s context retrieval and code correction mechanisms. The following analysis examines the specific design choices in the baselines that may contribute to this performance gap.

While \symprompt{} is designed to guide an LLM in generating tests for specific conditions specified within comments, we found its effectiveness in our experiments to be limited by the absence of code dependency information. We observed that its complex prompt structure, when not grounded with the necessary context, complicated the LLM's generation process. This suggests that the approach is robust but requires sufficient contextual input to steer the LLM effectively, and without it, it can struggle to match the performance of simpler methods.

The RAG-based approaches (\rag{}, \codeqa{}, and \draco{}) showed varied performance, highlighting different trade-offs in context retrieval. For instance, \draco{} is a sophisticated tool employing dedicated static analysis for Python. While effective, its reliance on import statements for dependency analysis means it can sometimes miss crucial intra-file context, such as helper functions or constants defined in the same file. On the other hand, \codeqa{} and \rag{} are designed to retrieve full code snippets of functions or classes, including intra-file context. This approach, however, sometimes led to an excess of retrieved context, particularly in larger projects, such as Tornado. This frequently caused the context to exceed the LLM's token limitation, resulting in the loss of potentially important information. Interestingly, our simpler baseline \rag{}, which retrieves a maximum of three code snippets, often performed better than the more complex RAG tools, suggesting that a more constrained and focused context can be more effective than a larger, unverified one.

\tool{}'s strong performance stems from two primary mechanisms. First, for higher coverage, \tool{} leverages LSP features like "go to definition" and "find references" to construct a precise semantic dependency graph. This provides the LLM with a complete yet concise context, avoiding the noise of generic RAG and the intra-file blind spots of other methods. Second, to achieve a higher valid rate, \tool{} utilizes real-time diagnostics on top of LSP to detect and correct syntactic errors in the generated code iteratively. This self-correction loop consistently improves the valid rate of generated test cases, as evidenced by the consistently high valid rates in Table~\ref{tab:main_results}. The specific contribution of these LSP-driven features is quantified in our ablation study (\S\ref{sec:ablation}).

The improvement in the valid rate for the Python projects (Black and Tornado) was less obvious than for the Java and Golang projects. This is an expected outcome, as Python is a dynamically typed language, which limits the language server's ability to catch all potential errors statically before execution. Conversely, for projects written in statically-typed languages like Java and Go, \tool{} consistently delivered significant improvements in validity rates because the language server could identify a broader range of potential errors.

We did not include Copilot in our experiments because it heavily relies on human interaction during the unit test generation. In detail, it requires manual clicks to trigger its functionality, which prevents large-scale automated experimentation as it does not provide automated APIs for programmatic code generation~\cite{stackoverflow2023githubcopilot,github2023discussion}. Furthermore, it assumes that users will perform context collection and error correction during interaction, introducing human factors that make fair and consistent evaluation difficult.

\subsection{Latency and Cost}

\begin{figure}[!ht]
  \centering
  \includegraphics[width=1\linewidth]{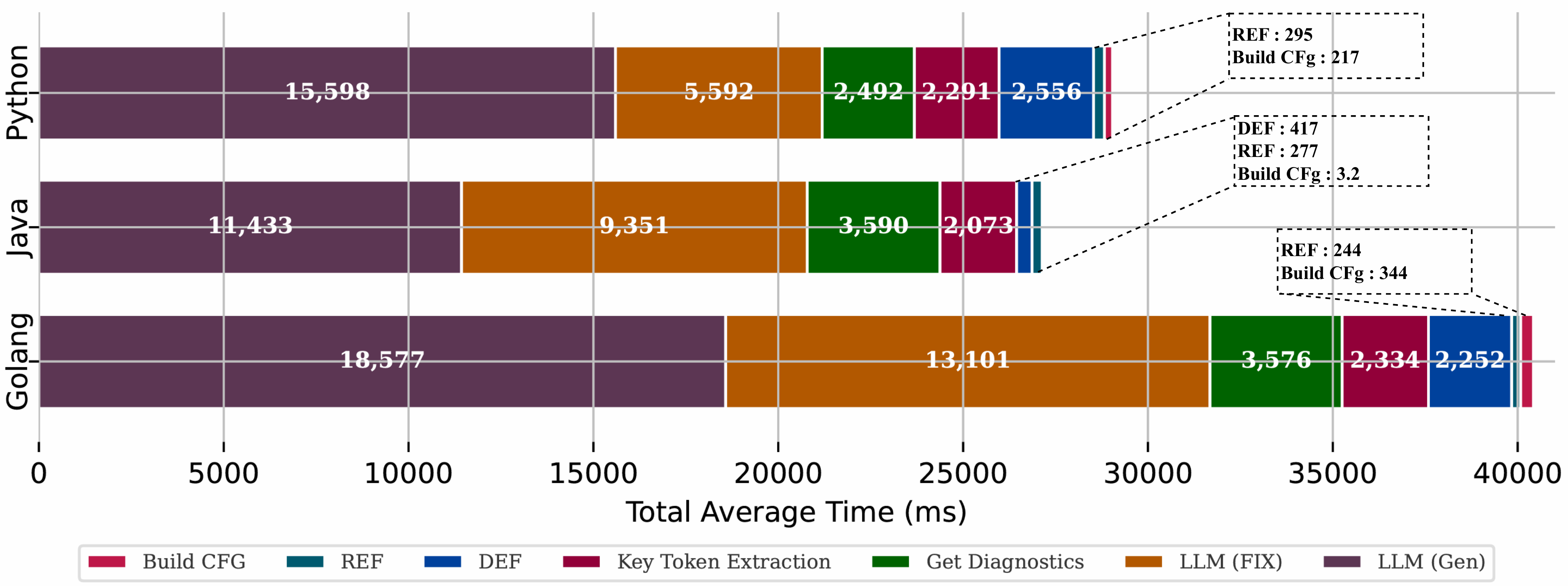}
  \caption{Test generation cost distribution per focal method, classified by programming languages.}
  \label{fig:cost}
\end{figure}

\begin{table}[h!]
    \centering
    \caption{Overall token usage distribution and comparison.}
    \label{tab:token_overhead_comparison}
    \resizebox{0.75\linewidth}{!}{ 
    \footnotesize
    \begin{tabular}{l|ccc|c}
    \toprule
    \textbf{Method} & \textbf{Java} & \textbf{Golang} & \textbf{Python} & \textbf{Averaged} \\
    \midrule
    \multicolumn{5}{l}{\textit{\tool{} - Breakdown}} \\
    \quad GEN    & 2,144 & 3,379 & 2,674 & 2,732 \\
    \quad FIX    & 3,252 & 1,458 & 586   & 1,832 \\
    \midrule
    \multicolumn{5}{l}{\textit{Total Token Usage Comparison}} \\
    \textbf{\tool{}} & 5,396 & 4,837 & 3,260 & 4,497 \\
    \codeqa       & 11,720 & 4,831 & 5,466 & 7,339 \\
    \rag     & 3,790  & 4,006 & 4,582 & 4,126 \\
    \draco        & --     & --    & 2,439 & --   \\
    \bottomrule
    \end{tabular}
    }
\end{table}
For a code generation tool to be practical in an interactive development workflow, it must operate with acceptable latency and cost. To answer RQ2, we assess whether \tool{} meets this real-time requirement by measuring its performance overhead in terms of latency and cost.
The evaluation was conducted on the projects from Table~\ref{tab:dataset-stats} using the \gptFourO{} API, following the same experimental setup as in \S\ref{sec:evrq1}. 
For each focal method, we measured the time and tokens consumed across all processes of the \tool{} pipeline, including a maximum of five self-correction attempts. The results of latency, averaged by programming language, are detailed in Figure~\ref{fig:cost}.

On average, our tool requires 28.27 seconds and 4,497 tokens per focal method to generate and refine a unit test. As detailed in ~\ref{fig:cost}, the majority of this time (approximately 70\%) is spent on LLM API querying for generation and refinement. Our retrieval strategy, which includes key token extraction and leveraging the LSP for reference and definition providers, accounts for about 5 seconds of the total time. While the refinement stage constitutes over 30\% of the generation time, we consider this a necessary and acceptable trade-off for the improvements it yields in both line coverage and the rate of valid tests.

As shown in Table~\ref{tab:token_overhead_comparison}, \tool{} averagely uses 2,732 tokens for the initial generation and 1,832 for fixing. While \tool{}'s token consumption is not the lowest in all cases, we consider the gains in coverage justify the trade-off.
For instance, compared to \codeqa{}, \tool{} uses 39\% fewer tokens while achieving a 135\% improvement in coverage. When measured against \draco{}, \tool{} uses 9.6\% more tokens for generation and 24\% more for fixing, but these increases yield coverage improvements of 30\% and an additional 6\%, respectively. Furthermore, \tool{} demonstrates its efficiency against a \rag{} approach, delivering a 174\% coverage improvement with an 8.9\% increase in total tokens used for generation and fixing.

\subsection{Ablation Study}
\label{sec:ablation}

\begin{table}[t]
    \centering
    \caption{Comparison of valid rate and line coverage across different versions of \tool{}. This shows that each component of \tool{} shows meaningful improvement both on coverage and valid rate.}
    \label{tab:ablation}
    \resizebox{\linewidth}{!}{
    \begin{tabular}{@{} cl *{3}{S[table-format=1.2]} *{3}{S[table-format=1.2]} @{}}
      \toprule
      \multirow{2.5}{*}{\textbf{Project}} & \multirow{2.5}{*}{\textbf{Model}} & \multicolumn{3}{c}{\textbf{Coverage}} & \multicolumn{3}{c}{\textbf{Valid Rate}} \\
      \cmidrule(lr){3-5} \cmidrule(lr){6-8}
      & & {\naive} & {\toolMinus} & {\tool{}} & {\naive} & {\toolMinus} & {\tool{}} \\
      \midrule
      \multirow{3}{*}{\black{}} 
      & \gptFourOmin{}     & 22.72\% & 39.51\% & 41.91\% & 57.32\% & 72.04\% & 79.73\% \\ 
      & \gptFourO{}        & 23.56\% & 42.30\% & 48.10\% & 56.58\% & 77.99\% & 85.15\% \\ 
      & \deepseek{}        & 39.74\% & 53.36\% & 56.89\% & 75.62\% & 87.02\% & 88.03\% \\ 
      \midrule 
      \multirow{3}{*}{\tornado{}}
      & \gptFourOmin{}     & 27.52\% & 56.72\% & 57.78\% & 67.14\% & 81.54\% & 81.84\% \\ 
      & \gptFourO{}        & 29.55\% & 60.03\% & 60.90\% & 54.36\% & 82.00\% & 84.53\% \\ 
      & \deepseek{}        & 46.04\% & 63.70\% & 64.62\% & 89.56\% & 90.36\% & 91.13\% \\ 
      \midrule
      \multirow{3}{*}{
      \parbox[c]{1cm}{\centering Commons \\ CLI}
      }
      & \gptFourOmin{}     & 04.59\% & 27.10\% & 33.29\% & 13.40\% & 23.49\% & 45.08\% \\ 
      & \gptFourO{}        & 12.71\% & 23.16\% & 34.69\% & 32.61\% & 28.59\% & 48.12\% \\ 
      & \deepseek{}        & 06.46\% & 28.77\% & 37.72\% & 17.20\% & 31.32\% & 60.52\% \\ 
      \midrule
      \multirow{3}{*}{
      \parbox[c]{1cm}{\centering Commons \\ CSV}
      }
      & \gptFourOmin{}     & 26.70\% & 69.77\% & 80.51\% & 15.74\% & 37.43\% & 82.85\% \\ 
      & \gptFourO{}        & 39.16\% & 76.05\% & 78.33\% & 35.69\% & 54.45\% & 90.9\% \\ 
      & \deepseek{}        & 32.67\% & 75.29\% & 83.20\% & 36.76\% & 49.29\% & 90.95\% \\ 
      \midrule
      \multirow{3}{*}{\cob}
      & \gptFourOmin{}     & 01.36\% & 09.91\% & 23.03\% & 01.19\% & 07.13\% & 23.88\% \\ 
      & \gptFourO{}        & 02.72\% & 12.56\% & 27.60\% & 01.79\% & 08.91\% & 33.27\% \\ 
      & \deepseek{}        & 11.58\% & 25.61\% & 37.22\% & 09.13\% & 21.78\% & 34.65\% \\ 
      \midrule
      \multirow{3}{*}{\logrus}
      & \gptFourOmin{}     & 02.32\% & 11.55\% & 23.71\% & 03.33\% & 18.86\% & 34.02\% \\ 
      & \gptFourO{}        & 00.65\% & 10.58\% & 27.75\% & 00.83\% & 15.00\% & 32.02\% \\ 
      & \deepseek{}        & 10.67\% & 13.55\% & 21.81\% & 22.50\% & 17.05\% & 33.11\% \\ 
      \bottomrule
    \end{tabular}
    }
  \end{table}

To answer RQ3, we conducted an ablation study to isolate and quantify the contribution of each of \tool{}'s core components. We evaluated three configurations, with same setup from \S\ref{sec:evrq1}:
\begin{itemize}[leftmargin=1em]
\item \textbf{\naive{}}: A baseline using the same prompt template as \tool{} but without LSP-guided context or code fixing.
\item \textbf{\toolMinus{}}: The baseline enhanced with our context retrieval on top of LSP features.
\item \textbf{\tool{}}: The complete \tool{} system, incorporating both context retrieval and the real-time code fixing module.
\end{itemize}

The results, averaged over five runs and presented in Table~\ref{tab:ablation}, reveal two key findings.
First, context retrieval is crucial for generating comprehensive tests. As shown in Table~\ref{tab:ablation}, adding context (\toolMinus{}) substantially increases line coverage over the \naive{} baseline across all projects and LLMs. Specifically, \toolMinus{} improved line coverage by 26\% to 1528\% for Golang, 82\% to 490\% for Java, and 34\% to 106\% for Python. 
This demonstrates that the precise, semantically aware context provided by our context retrieval on top of the LSP module enables the LLM to understand class structures, method dependencies, and necessary initializations.

Second, the real-time fixing module is vital for maintaining correctness, especially when the context is complex. While rich context boosts coverage, it can be a double-edged sword. For instance, with a complex project like \cli, the large volume of context can distract the LLM, leading to syntactically incorrect code and a lower valid rate for \toolMinus{}. The full \tool{} system mitigates this issue. By adding the fixing module, \tool{} improved line coverage by an additional 45\% to 162\% for Golang, 3\% to 49\% for Java, and 1\% to 13\% for Python over \toolMinus{}. More importantly, it increased the valid rate by 59\% to 273\% for Golang, 67\% to 121\% for Java, and 0.5\% to 10\% for Python. The improvements for Python are less pronounced, which is an expected outcome due to the language's dynamic nature. Python's dynamic nature means that its LSP diagnostics, which rely on static analysis, cannot detect certain error classes that would cause compile-time failures in statically-typed languages.
Nevertheless, the fixing module proves its general utility by consistently enhancing both the validity and coverage of generated tests across all three languages, demonstrating that real-time correction is a crucial component for robust test generation.
\section{Discussion}

{\bf Fault-Finding Capability.}
While \tool{} is designed to maximize test coverage, it does not explicitly optimize for the quality of test assertions. Since effective assertions are critical for detecting faults, we conducted an experiment to evaluate this capability. We injected faults into the source code using standard mutation tools (Pit~\cite{Coles2016PIT} and MutPy~\cite{Derezinska2015MutPy}) and measured whether the generated tests could detect them.
\begin{table}[h!]
    \centering
    \small
    \caption{Comparison of mutation scores across all baselines.}
    \label{tab:fault-finding}
    \resizebox{\linewidth}{!}{
    \begin{tabular}{ll|ccccc}
        \toprule
        \textbf{Project} & \textbf{Model} & \textbf{\tool{}} & SymPrompt & CodeQA & StandardRAG & DraCo \\
        \midrule
        \multirow{3}{*}{\black} 
            & \deepseek{}  & \textbf{25.21\%} & 12.75\% & 11.85\% & 5.03\% & 7.73\% \\
            & \gptFourO{}      & \textbf{30.28\%} & 21.04\% & 18.47\% & 9.69\% & 17.03\% \\
            & \gptFourOmin{}  & \textbf{31.90\%} & 15.01\% & 8.18\%  & 2.31\% & 8.71\% \\
        \midrule
        \multirow{3}{*}{\tornado} 
            & \deepseek{}  & \textbf{35.39\%} & 1.46\%  & 2.03\%  & 21.16\% & 2.03\% \\
            & \gptFourO{}      & \textbf{30.37\%} & 9.11\%  & 7.66\%  & 18.16\% & 21.69\% \\
            & \gptFourOmin{}  & \textbf{24.96\%} & 5.15\%  & 13.62\% & 9.28\%  & 15.68\% \\
        \midrule
        \multirow{3}{*}{
            \parbox[c]{0.85cm}{\centering Commons \\ CLI}
            }
            & \deepseek{}  & \textbf{19.67\%} & 3.22\%  & 4.21\%  & 12.73\% & \textemdash{} \\
            & \gptFourO{}      & \textbf{19.01\%} & 1.32\%  & 7.19\%  & 0.41\%  & \textemdash{} \\
            & \gptFourOmin{}  & \textbf{19.92\%} & 0.08\%  & 1.40\%  & 0.74\%  & \textemdash{} \\
        \midrule
        \multirow{3}{*}{
            \parbox[c]{0.85cm}{\centering Commons \\ CSV}
            }
            & \deepseek{}  & \textbf{58.50\%} & 11.75\% & 39.63\% & 17.63\% & \textemdash{} \\
            & \gptFourO{}      & \textbf{50.50\%} & 0.25\%  & 9.38\%  & 5.63\%  & \textemdash{} \\
            & \gptFourOmin{}  & \textbf{50.13\%} & 1.13\%  & 7.63\%  & 2.75\%  & \textemdash{} \\
        \bottomrule
    \end{tabular}
    }
\end{table}
As summarized in Table~\ref{tab:fault-finding}, \tool{} consistently outperforms the baselines, suggesting that its accurate context retrieval not only boosts coverage but also enhances fault detection.

{\bf Comparison with SBST Tool.}
To contextualize the performance of our \tool{}, we conducted a direct comparison with \evosuite{}, a widely-used SBST tool for java. For fair comparison, \evosuite{} was configured to generate unit tests per method under a fixed time budget of 28 seconds per method under test. The results are summarized in Table~\ref{tab:sbst-comparison}.
\begin{table}[h!]
    \centering
    \caption{Comparison with EvoSuite on Java projects.}
    \label{tab:sbst-comparison}
    \begin{tabular}{l|cc|cc}
    \toprule
    \multirow{2}{*}{\textbf{Project}} & \multicolumn{2}{c|}{\textbf{Coverage}} & \multicolumn{2}{c}{\textbf{Valid Rate}} \\ \cline{2-5} 
     & \tool{} & \evosuite & \tool{} & \evosuite \\ \midrule
    \cli & 37.72\% & \textbf{56.33\%} & 60.52\% & \textbf{100\%} \\
    \csv & \textbf{83.20\%} & 69.70\% & 90.95\% & \textbf{100\%} \\ \bottomrule
    \end{tabular}
\end{table}
The results indicate that \tool{} achieves competitive code coverage against \evosuite{}. \evosuite{} consistently achieves a 100\% valid generation rate. This is an expected outcome, as its test generation process is rooted in a strict code mutation strategy that inherently guarantees syntactically correct and compilable tests. In contrast, \tool{} is designed with a different primary goal: to support multiple languages through a lightweight static analysis framework. This design choice involves a trade-off, where we sacrifice a guaranteed valid rate for broader applicability and scalability. Nevertheless, the result demonstrates that \tool{} still delivers coverage results that are comparable to \evosuite{}. 

{\bf Threats to Validity.}
We identify two primary external threats to the generalizability of our approach.
First, the effectiveness of \tool{} is inherently coupled with the correctness of the underlying LSP server implementation for a given language. 
Our approach relies on the LSP server to provide accurate static analysis results (e.g., definitions, references, and type hierarchies). Consequently, if a programming language lacks robust LSP support, or for servers with erratic or incomplete analysis capabilities, the quality of the retreived context would be degraded, potentially impacting the quality of the generated code. 
Second, the behavior of \tool{} can be influenced by environment-specific configurations. 
For example, the behavior of LSP server is affected by different IDEs.
We observed that for a Java project, VS Code~\cite{vscode} might define the class path as the project's top-level directory, whereas Cursor~\cite{cursoride} sets it to the \texttt{src/} subdirectory.
Such discrepancies can affect which files are analyzed, potentially affecting the LSP's static analysis results. Our prototype accommodates these variations through configuration.

{\bf Extensibility to Other Languages.}
One of the key design goals of \tool{} is extensibility. Its architecture is modular, comprising: (1) a language-specific module for AST parsing and key feature extraction, and (2) language-agnostic modules that leverage the LSP for context retrieval and code fixing.
To adapt \tool{} for a new programming language, manual implementation is only necessary for the first module. This task involves translating the language's specific tree-sitter based AST nodes into \tool{}'s IRs. 
This is a well-defined task requiring a modest implementation effort. 
Integrating the LSP-based capabilities is straightforward. It only requires the installation of the appropriate language server (e.g., as an IDE extension). Once installed, \tool{} can immediately utilize the server's rich functionalities, such as ``Go To Definition,'' without any further modification to its core logic.
\section{Related Work}
\label{sec:related-work}


{\bf Existing Approaches on Unit Test Generation.}
Automated unit test generation has been a long-standing goal in software engineering, with two primary waves of innovation: classical search-based and symbolic techniques, and more recent LLM-based approaches.
Classical techniques primarily focus on maximizing a specific criterion, most often code coverage. Search-Based Software Testing (SBST) tools~\cite{UTGen,fraser2011evosuite,lin2020evosuiteplusplus1,pacheco2007randoop, sapozhnikov2024testspark} for Java, and tools~\cite{lukasczyk2022pynguin, lemieux2023codamosa, altmayer2024coverup} for Python, employ evolutionary algorithms or random testing to generate test suites. While powerful, these methods are \textit{inappropriate for real-time}. They rely on whole-program compilation, heavy instrumentation, and numerous execution iterations---steps that are too slow and fragile when the codebase is in flux, changing every few minutes during active development. Their operation assumes a stable, buildable project state, which is often not the case in the early, exploratory stages of writing new code.
More recently, LLMs have been applied to test generation, often outperforming classical tools in terms of the readability and initial quality of generated tests~\cite{xie2023chatunitest, 2024mutap,wang2024chat,altmayer2024coverup,symprompt}. However, most of these works have selected an iterative loop either for error detection or coverage feedback. For instance, Altmayer et al.~\cite{altmayer2024coverup} propose coverage-guided pipelines where an LLM generates tests, which are then executed to gather coverage data; this feedback is used to prompt the LLM to cover the remaining branches. 
While these approaches have improved previous work in terms of latency and coverage, they are still inappropriate for real-time scenarios.
Furthermore, both classical and many LLM-based approaches are tailored for specific programming languages (e.g., Java and Python), limiting their generalizability.

{\bf Retrieval-Augmented Generation.}
RAG~\cite{lewis2020retrieval,karpukhin2020dense} mitigates LLM's limitation of having finite training data. Initially proposed for NLP, it has been widely applied to code, notably through repository-level techniques. These use code-specific features for context retrieval, such as AST-based chunking~\cite{code_qa, ding2023coderag, shou2023codegpt} or program-analysis-based retrieval~\cite{zhang2023repocoder,liu2024codexgraph, liu2024graphcoder, chen2022codet}. 
Repository-level RAG techniques are representative of this effort.
They used code-specific features to retrieve context, such as AST-based chunking~\cite{code_qa, ding2023coderag, shou2023codegpt}, or program-analysis based retrieval~\cite {zhang2023repocoder,liu2024codexgraph, liu2024graphcoder, chen2022codet}.
These techniques expose a fundamental trade-off. AST-based chunking is multi-lingual, but its reliance on superficial features leads to imprecise retrieval. In contrast, program-analysis-based retrieval achieves far greater precision by understanding the code's execution and data-flow relationships. However, this precision comes at a steep price: they require building and maintaining complex, language-specific static analyzers, thereby limiting their applicability to one specific language~\cite{DraCo,liu2024graphcoder,liu2024codexgraph}. We compare \tool{} with \draco{}\cite{DraCo} and \codeqa{}\cite{code_qa} as they are the only open-sourced, referred code-aware RAG works.

{\bf Neuro-Symbolic Approaches in Software Engineering.}
Recent neuro-symbolic approaches combine LLMs with program analysis for enhanced code intelligence. This synergy aids code auditing and bug detection, with tools like LLMDFA~\cite{wang2024llmdfa} and DeepConstr~\cite{deepconstr} using LLMs for dataflow and constraint generation while symbolic techniques validate paths and mitigate hallucinations. For unit test generation, Tratto~\cite{daka2024tratto} uses a neural module to propose tokens and a symbolic module to constrain the search space based on program context.
However, a common limitation is that they are tailored to a single programming language. In contrast, \tool{} addresses this gap by leveraging the LSP to perform language-agnostic neuro-symbolic unit test generation.
\section{Conclusion}
We addressed the challenge of generating high-coverage unit tests for modern, multi-language software systems in real-time. We identified a key deficiency in existing approaches: their inability to precisely retrieve the necessary context, which hinders the generation of high-coverage and valid tests. To overcome this, we introduced \tool{}, a novel framework that leverages the LSP to obtain precise code context. \tool{} incorporates a hybrid analysis strategy to distill essential, branch-governing symbols from the retrieved context and a compile-free self-repair mechanism to ensure the syntactic validity of the generated tests without the overhead of compilation. Our extensive evaluation on real-world projects in Java, Python, and Golang demonstrates that \tool{} significantly improves both line coverage and the rate of valid test generation across different programming languages and underlying LLMs.

\section*{Acknowledgements}
We appreciate the reviewers' valuable and insightful comments.
This research is partly sponsored by CCF-Tencent Rhino-Bird Fund Program (No. 20242001274) and NSFC Program (No. 62525207).

\newpage
\bibliographystyle{ACM-Reference-Format}
\bibliography{ref.bib}

\end{document}